\documentclass[acmsmall]{acmart}
\usepackage[T1]{fontenc}
\usepackage{adjustbox}
\usepackage{threeparttable}
\usepackage{multirow}
\usepackage{tabularx}
\usepackage{enumitem}
\usepackage{pifont}
\usepackage{caption}
\usepackage{graphicx}
\usepackage{float} 
\usepackage{subcaption}
\usepackage{makecell}
\usepackage{soul}
\usepackage{tcolorbox}
\usepackage{mdframed}
\tcbuselibrary{skins}

\usepackage{caption}
\usepackage{graphicx}
\usepackage{float} 
\usepackage{subcaption}
\usepackage{soul}
\usepackage{mdframed}
\usepackage{makecell}

\newcommand{\tech}{RevAgent}

\AtBeginDocument{%
  }

\setcopyright{acmlicensed}
\copyrightyear{2025}
\acmYear{2025}
\acmDOI{XXXXXXX.XXXXXXX}
\acmConference[Conference acronym 'XX]{Make sure to enter the correct
  conference title from your rights confirmation email}{June 03--05,
  2018}{Woodstock, NY}
\acmISBN{978-1-4503-XXXX-X/2018/06}
\newcounter{finding}
\newcommand{\finding}[1]{\refstepcounter{finding}
    \vspace{4pt}
    \begin{mdframed}[linecolor=gray,roundcorner=12pt,backgroundcolor=gray!15,linewidth=3pt,innerleftmargin=2pt, leftmargin=0cm,rightmargin=0cm,topline=false,bottomline=false,rightline = false]
        #1
    \end{mdframed}
    \vspace{4pt}
}

\newcommand{\prompt}[1]{
\begin{tcolorbox}[colback=yellow!5!white, size=small, boxrule=0pt, frame hidden]
#1
\end{tcolorbox}
}

\begin{document}

\title{Issue-Oriented Agent-Based Framework for Automated Review Comment Generation}

\author{Shuochuan Li}
\email{lishuochuan@tju.edu.cn}
\affiliation{%
  \institution{Tianjin University}
  \city{Tianjin}
  \country{China}
}

\author{Dong Wang}
\authornote{Corresponding Author.}
\email{dong_w@tju.edu.cn}
\affiliation{%
  \institution{Tianjin University}
  \city{Tianjin}
  \country{China}
}

\author{Patanamon Thongtanunam}
\affiliation{%
  \institution{The University of Melbourne}
  \city{Victoria}
  \country{Australia}
}

\author{Zan Wang}
\affiliation{%
  \institution{Tianjin University}
  \city{Tianjin}
  \country{China}
}
\author{Jiuqiao Yu}
\affiliation{%
  \institution{University of California Berkeley}
  \city{Berkeley}
  \state{California}
  \country{USA}
}

\author{Junjie Chen}
\affiliation{%
  \institution{Tianjin University}
  \city{Tianjin}
  \country{China}
}

\renewcommand{\shortauthors}{Li et al.}

\begin{abstract}
Code review (CR) is a crucial practice for ensuring software quality. Various automated review comment generation techniques have been proposed to streamline the labor-intensive process. However, existing approaches heavily rely on a single model to identify various issues within the code, limiting the model's ability to handle the diverse, issue-specific nature of code changes and leading to non-informative comments, especially in complex scenarios such as bug fixes. To address these limitations, we propose RevAgent, a novel agent-based issue-oriented framework, decomposes the task into three stages: (1) Generation Stage, where five category-specific commentator agents analyze code changes from distinct issue perspectives and generate candidate comments; (2) Discrimination Stage, where a critic agent selects the most appropriate issue-comment pair; and (3) Training Stage, where all agents are fine-tuned on curated, category-specific data to enhance task specialization. Evaluation results show that RevAgent significantly outperforms state-of-the-art PLM- and LLM-based baselines, with improvements of 12.90\%, 10.87\%, 6.32\%, and 8.57\% on BLEU, ROUGE-L, METEOR, and SBERT, respectively. It also achieves relatively higher accuracy in issue-category identification, particularly for challenging scenarios.  Human evaluations further validate the practicality of RevAgent in generating accurate, readable, and context-aware review comments. Moreover, RevAgent delivers a favorable trade-off between performance and efficiency.
\end{abstract}




\maketitle

\section{Introduction}

Code review (CR) is a fundamental practice in software quality assurance, serving to uncover defects, suboptimal design choices, and potential bugs while safeguarding the long‑term quality and maintainability of a codebase~\cite{bacchelli2013expectations, wang2021can, wang2021understanding}. 
Despite its benefits in detecting issues early, promoting knowledge sharing, and enforcing coding standards, code review is labor-intensive, time-consuming, and highly dependent on reviewers’ expertise~\cite{macleod2017code}. 
To address these challenges, automated code review approaches have gained increasing attention as a means to enhance efficiency and consistency~\cite{DBLP:journals/tse/TufanoDMCB24}.
A central component of such automation is review comment generation, which involves detecting issues in code changes and providing actionable suggestions. 
The quality of these comments is essential, as they directly influence the effectiveness of subsequent code refinements.
Therefore, generating accurate and relevant review comments is critical to ensuring that issues are thoroughly addressed and properly documented.

Recently, increasing efforts have been devoted to automating review comment generation, resulting in a variety of techniques~\cite{DBLP:conf/icse/TufanoPTPB21,DBLP:conf/icse/TufanoMMPPB22,DBLP:conf/sigsoft/LiLGDJJMGSFS22,DBLP:journals/pacmse/SghaierS24,DBLP:conf/issre/LuYLYZ23,DBLP:journals/tosem/YuRSZSWWXW25,DBLP:journals/tse/NashaatM24, DBLP:conf/sigsoft/HongTTA22}, including approaches based on information retrieval (IR), pre-trained language models (PLMs), and large language models (LLMs).
For instance, Li et al.~\cite{DBLP:conf/sigsoft/LiLGDJJMGSFS22} proposed CodeReviewer, leveraging CodeT5 for pre-training on large-scale code review datasets.
To address the limited generalizability of pre-trained models, recent studies have increasingly adopted LLMs with task-specific fine-tuning.
For instance, Lu et al.~\cite{DBLP:conf/issre/LuYLYZ23} introduced LLaMA-Reviewer by fine-tuning a base LLaMa model, while Nashaat et al.~\cite{DBLP:journals/tse/NashaatM24} and Yu et al.~\cite{DBLP:journals/tosem/YuRSZSWWXW25} fine-tuned LLMs on a benchmark dataset for specific requirements of code review, all of them showing the promise.
These efforts collectively demonstrate the promise of LLM-based approaches in enhancing the quality and relevance of generated review comments.

However, existing approaches typically rely on a single model to analyze code and detect diverse issue types. 
This lack of specialization blurs focus and often yields superficial, non-informative comments, undermining review effectiveness.
As noted by Tufano et al.~\cite{DBLP:journals/tse/TufanoDMCB24}, such methods also overlook the heterogeneity of code-change issues.
Specifically, well-established PLM- and IR-based approaches primarily skew toward easy refactorings (e.g., simple variable or constant changes) and perform poorly on more critical scenarios, like bug fixes, testing, and logging.
A major driver of this disparity is dataset bias: complex issue types are underrepresented in existing benchmarks, pushing models toward frequent, simpler patterns and leaving them without the issue-specific knowledge or contextual sensitivity required to generate meaningful comments for high-impact cases.
Recent industrial research~\cite{goldman2025typescodereviewcomments} further emphasizes that LLM-based reviewers should balance comment types while improving clarity and relevance.

On the other hand, while LLM-based approaches have exhibited superior performance in terms of textual similarity metrics (e.g., BLEU), their capability to generate issue-specific and contextually relevant comments remains largely unexplored. 
Tufano et al.~\cite{DBLP:journals/tse/TufanoDMCB24} made an initial attempt to address this limitation by introducing a prompt-based approach that guides an LLM using a taxonomy of five root code-related categories.
However, their evaluation revealed that even with such prompts, a single LLM still struggled to grasp the nuanced logic of each issue category
on the code-to-comment generation task, underscoring the need for more structured and context-sensitive strategies.
Therefore, we argue that leveraging diverse comment generation experts, each focusing on a distinct issue category,  could overcome the limitations of one-size-fits-all approaches and produce more accurate, issue-oriented review comments for practitioners.

To enable LLMs to better focus and capture a broader range of issues, we propose \textbf{\tech{}}, a novel multi-issue-oriented framework designed to enhance the diversity and depth of generated review comments.
To effectively handle different issue types, \tech{} leverages an agent-based architecture that transforms the traditional single-LLM pipeline into a collaborative multi-agent system, each specialized in a specific issue type.
Inspired by the recent success of agent-based systems for natural-language-to-code tasks (e.g., code generation~\cite{DBLP:conf/acl/ZhangLLSJ24, DBLP:conf/acl/IslamAP24}), this work adopts a multi-agent system for this code-to-natural-language task.    
Rather than designing a one-stage review comment generation pipeline, \tech{} has a three-stage process (i.e., the \textit{Generation Stage}, \textit{Discrimination Stage}, and \textit{Training Stage}), to enhance performance through specialization and targeted augmentation.
\tech{} addresses a central challenge in identifying the appropriate perspective for a given code change, especially in the absence of explicitly labeled issue categories.
In particular, \tech{} introduces three coordinated stages:
(1) \textit{Generation Stage}, where five category-specific commentator agents act as specialized reviewers. Each agent analyzes the code change from a distinct issue-category perspective (e.g., refactoring, bug-fixing, and logging) and produces a corresponding candidate review comment.
(2) \textit{Discrimination Stage}, where a critic agent evaluates the candidate comments, infers the most relevant issue category, and selects the most appropriate final comment.
To equip each agent involved in the aforementioned stages with the necessary expertise, \tech{} incorporates a foundational stage: (3) \textit{Training Stage}, where each commentator agent is fine-tuned on data specifically curated for its assigned issue category.
The critic agent, responsible for cross-category evaluation, is trained on a diversified dataset in which similar code changes are paired with review comments from multiple issue perspectives.

To evaluate \tech{}, we conducted extensive experiments on the Curev~\cite{DBLP:journals/corr/abs-2502-03425}, which contains 20,000 review comment instances categorized into five issue types: Refactoring, Bugfix, Testing, Logging, and Documentation.
The results demonstrate that \tech{} outperforms both PLM-based and LLM-based state-of-the-art baselines, achieving average improvements of 12.90\%, 10.87\%, 6.32\%, and 8.57\% in BLEU, ROUGE-L, METEOR, and SBERT scores, respectively.
In terms of issue-category identification, \tech{} achieves an overall accuracy of 60.20\%, including 21.69\% accuracy in the challenging Bugfix category, where existing approaches typically perform below 5\%.
To assess the contributions of individual components, we performed ablation experiments by replacing the category-specific commentator agents and the critic agent with alternative implementations.
Results show that using multiple specialized commentator agents significantly improves the generation of category-aligned comments compared to a single unified agent, and the critic agent plays the most crucial role in accurately identifying issue categories.
Furthermore, human evaluations on a total of 1,920 samples, with 384 samples evaluated for each of the five approaches, confirm that the review comments generated by \tech{} are indeed helpful for developers, offering improved readability, accuracy, and alignment with the intended issue category.
From an efficiency standpoint, although the agent-based design introduces a modest additional latency of 0.038 seconds per generation compared to a single LLM, the performance gains justify this trade-off and remain within a practically acceptable range.
Finally, we manually analyze the root causes of low-quality generated comments to guide future research, revealing that 48\% of failure cases are attributed to the lack of project-specific knowledge (e.g., coding standards).


\smallskip
\noindent
\textbf{Contributions.} 
To summarize, this paper makes the following key contributions:
\begin{itemize}
    \item We propose a novel agent-based framework for automated review comment generation that enhances issue-oriented feedback by equipping category-specific experts and coordinating their reasoning to handle diverse issue types.
    \item We conduct a comprehensive evaluation of our framework on a large-scale dataset, demonstrating that \tech{} significantly outperforms state-of-the-art approaches in both quantitative metrics and qualitative assessments.
    \item We extensively evaluate various advanced open-source LLMs within our framework, examining models of different sizes to assess their effectiveness and scalability within our framework.
    \item To support future research, we publicly release our replication package~\cite{our/url}, which includes all necessary code and resources to reproduce the results.
\end{itemize}

\smallskip
\noindent
\textbf{Paper Organization.} The remainder of this paper is organized as follows. 
Section~\ref{sec:background} positions our work in relation to prior research.
Section~\ref{sec:methodology} details our proposed agent-based framework.
Section~\ref{sec:exmperiment} describes the experimental setup.
Section~\ref{sec:results} presents results and answers the research questions.
Section~\ref{sec:discussion} further examines efficiency and provides qualitative insights. 
Section~\ref{sec:threat} outlines threats to validity.
Finally, we conclude the paper and highlight directions for future work in Section~\ref{sec:conclusion}.

\section{Related Work}
\label{sec:background}
This section surveys related work on automated code review and agent-based AI to situate our research within existing literature.

\textbf{Automated Code Review.}
Modern Code Review (MCR) is a collaborative and lightweight practice widely adopted in contemporary and open-source software development, serving as a cornerstone of quality assurance and a vital mechanism for knowledge sharing among developers~\cite{bacchelli2013expectations, rigby2013convergent}.
The process begins when a developer submits a proposed code change, typically in the form of a pull request or merge request, through a version control platform (e.g., GitHub).
Reviewers are then assigned to examine the change, leave inline or general comments, and often engage in iterative discussions with the author to clarify intentions or suggest improvements. 
The author responds by revising the code, addressing feedback, or justifying design decisions. This cycle continues until the reviewers are satisfied with the change’s correctness, clarity, and overall quality. Once approved and after passing all automated checks, the change is merged into the main codebase.

To alleviate the substantial manual effort involved in code review, automation has garnered increasing attention in recent years.
Early efforts, such as CodeReviewer\cite{DBLP:conf/sigsoft/LiLGDJJMGSFS22} and AUGER\cite{DBLP:conf/sigsoft/LiYJYLHLZ22}, applied deep learning techniques based on PLMs to support review activities.
With the emergence of LLMs, more recent studies have explored their use in code review, improving both generation quality and evaluation methods.
For instance, LLaMA-Reviewer~\cite{DBLP:conf/issre/LuYLYZ23} fine-tuned the LLaMA model specifically for review tasks.
Building on this direction, subsequent work has investigated dataset augmentation, fine-tuning strategies, prompt engineering, and the design of collaborative AI systems~\cite{DBLP:conf/icse/GuoCXLL0024,DBLP:journals/tse/NashaatM24,DBLP:journals/tosem/YuRSZSWWXW25,DBLP:conf/emnlp/TangKSLLETKB24}.
Specifically, Nashaat et al.~\cite{DBLP:journals/tse/NashaatM24} and Yu et al.~\cite{DBLP:journals/tosem/YuRSZSWWXW25} constructed benchmark datasets to support LLM fine-tuning in review settings.
CodeAgent~\cite{DBLP:conf/emnlp/TangKSLLETKB24} introduced a multi-agent framework that simulates collaborative review dynamics through role-specific agents.
BitsAI-CR~\cite{DBLP:journals/corr/abs-2501-15134} adopted a two-stage architecture combining a RuleChecker and a ReviewFilter to enhance the precision of generated comments. 
More recently, \citet{DBLP:journals/ase/LiWWHWLH25} introduced CodeDoctor that generates multi-category review comments at one shot using one LLM.
However, this approach departs from the established paradigm of automated comment generation and may impose additional cognitive burden on practitioners.

Despite these advances, it is noted that most existing methods rely on single-model architectures, limiting their ability to handle diverse issue types and resulting in less precise, context-aware comments.
To address this limitation, we follow the established paradigm (for a given code change, generate an issue-specific review comment) and propose an agent-based framework that generates review comments from multiple perspectives and evaluates trade-offs to select the most relevant and informative as the final output.

\begin{figure*}[t]
  \centering
  \includegraphics[width=1\linewidth]{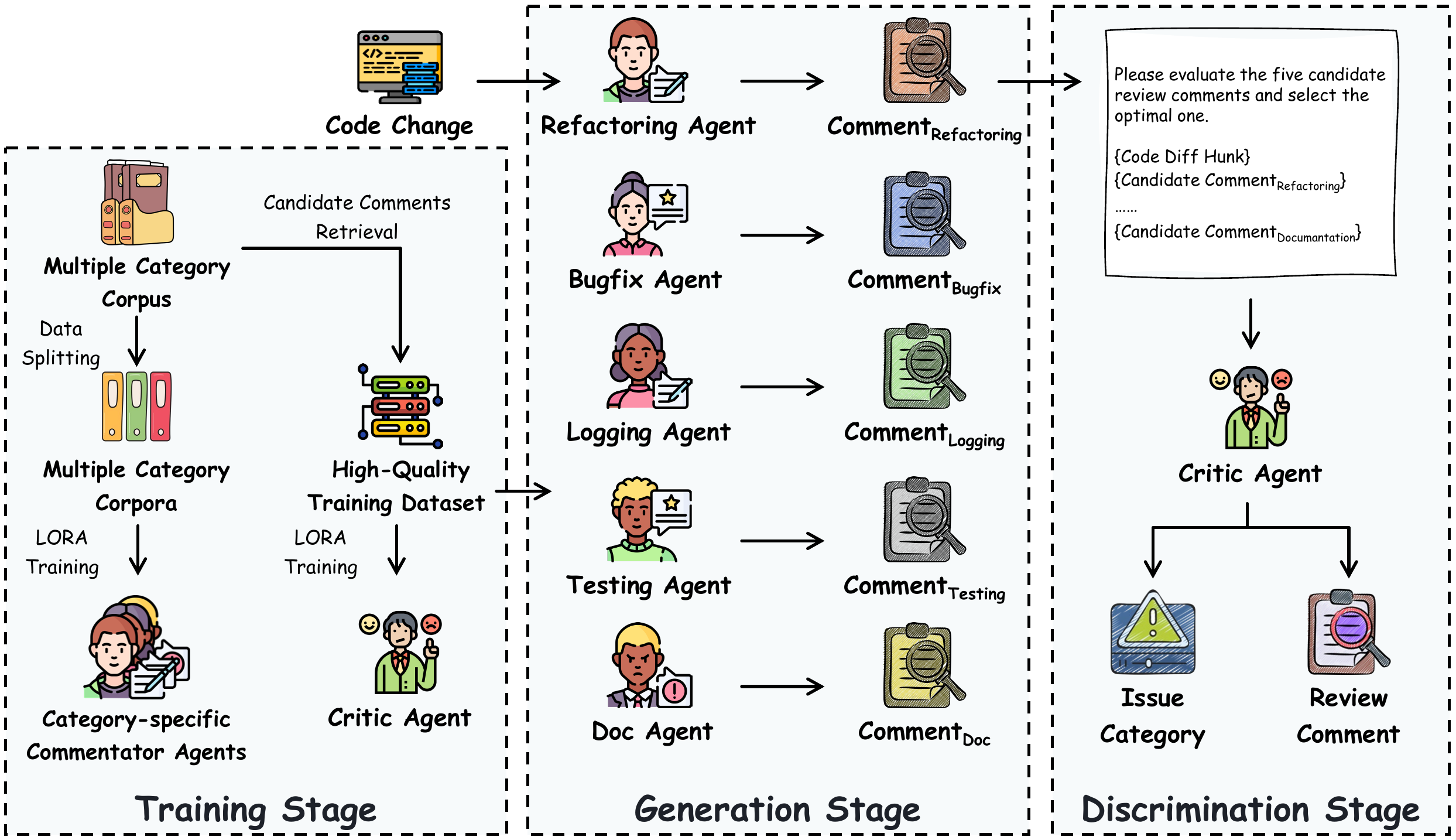}
  \caption{The overview of \tech{}} 
  \label{fig:overview}
\end{figure*}

\textbf{Agent-Based AI.}
It refers to artificial intelligence systems designed to achieve shared goals with humans or other AI systems.
Previous research has extensively explored the use of multiple LLMs in agent-based or multi-agent settings~\cite{DBLP:journals/corr/abs-2306-03314,DBLP:conf/acl/QianLLCDL0CSCXL24}. 
These approaches leverage inter-agent interactions to collectively enhance LLM capabilities, leading to improved overall performance across various scenarios~\cite{wei2023multipartychatconversationalagents,DBLP:journals/corr/abs-2303-17760,DBLP:conf/uist/ParkOCMLB23}.
For example, Akata et al.~\cite{DBLP:journals/corr/abs-2305-16867} examined LLM cooperation through orchestrated repeated games, while Cai et al.~\cite{DBLP:conf/iclr/Cai00CZ24} proposed an agent-based framework where large models act as tool-makers and small models as tool-users to reduce computational costs.

In the software engineering domain, agent-based AI is increasingly promising for a variety of scenarios~\cite{he2025llm}.
Li et al.\cite{DBLP:journals/corr/abs-2303-17760} and Qian et al.\cite{DBLP:conf/acl/QianLLCDL0CSCXL24} introduced multi-agent frameworks for software development, relying on natural language conversations rather than formalized software engineering artifacts.
Elicitation~\cite{DBLP:journals/corr/abs-2404-16045} simulates the requirement elicitation phase by assigning distinct personas to different agents for interactive role-play, while MapCoder~\cite{DBLP:conf/acl/IslamAP24} emulates the code generation cycle through four specialized LLM agents responsible for recalling relevant examples, planning, generating code, and debugging.
\citet{wang2025m2cvd} introduces the multi-model collaborative vulnerability detection
approach that leverages the strong capability of analyzing vulnerability semantics from LLMs. 
Their results show clear gains over single-model detectors.

Despite recent progress, applying agent-based AI to code review poses distinct challenges.
Unlike generic generation tasks, reviewers must interpret fine-grained diffs, infer underlying issue type (e.g., refactoring, bug fix, testing), and produce context-aware feedback.
A single, monolithic model tends to blur these distinctions and produce generic, low-value comments.
Building on these insights, we extend agent-based AI to automated code review with an agent-based framework tailored for issue-oriented comment generation, coupling category-specialist commentator agents with a critic that coordinates and selects the most relevant review comment.
\section{\tech{}}
\label{sec:methodology}

This section introduces \tech{}, a novel approach to generating review comments based on an issue-oriented agent-based system. 
As shown in Figure 1, the framework comprises three stages: \textit{Generation}, \textit{Discrimination}, and \textit{Training}. 
In the \textit{Generation Stage}, given a code change, \tech{} analyzes it from five distinct issue-category perspectives and generates corresponding candidate review comments. 
In the \textit{Discrimination Stage}, a critic agent evaluates these candidates based on factors such as correctness and severity, and selects the most appropriate issue-comment pair as the final output. 
To enhance the performance of both the category-specific commentator agents and the critic agent, the \textit{Training Stage} involves fine-tuning each agent on task-specific datasets. 


\subsection{Generation Stage}

The goal of this stage is to identify the primary issues in a given code change from diverse issue-category perspectives and generate corresponding candidate review comments.
To accomplish this, we deploy five category-specific commentator agents, each functioning as an independent code reviewer specialized in a particular issue type.
Each agent analyzes the code change to detect issues relevant to its assigned category and generates a potential suggestion for improvement.
These agents operate in a competitive setting, with each striving to produce the most accurate and contextually appropriate review comment from its respective perspective.

\begin{table}[]
\centering
\caption{The taxonomy of issue categories\label{tab:taxonomy} }

 \begin{adjustbox}{width=.6\linewidth,center}
 \begin{threeparttable}
\begin{tabular}{ll}
\toprule
\textbf{Category} & \textbf{Definition} \\
    \midrule
    Refactoring  & \makecell[l]{Suggestions to improve code structure.}  \\
    Bugfix   & \makecell[l]{Identifies and suggests fixes for bugs.}  \\
    Testing & \makecell[l]{Comments related to test cases.}  \\
    Logging & \makecell[l]{Suggestion for logging practices.}  \\
    Documentation & \makecell[l]{Recommendations for documentation changes.}  \\
    Others & \makecell[l]{Unspecified or ambiguous comments.} \\
\bottomrule
\end{tabular}
\end{threeparttable}
\end{adjustbox}
\end{table}

The prompt for each category-specific commentator agent contains two main components: a task description and a query code diff hunk.
In the task description section, we first define the agent’s role and specify the potential issue it should address, as highlighted in red in the example prompt.
Following Tufano et al.~\cite{DBLP:journals/tse/TufanoDMCB24}, issues are categorized into six major groups: \textit{Refactoring}, \textit{Bugfix}, \textit{Testing}, \textit{Logging}, \textit{Documentation}, and \textit{Others}, as summarized in Table\ref{tab:taxonomy}.
In line with prior research~\cite{DBLP:conf/icse/MuCSWW23, DBLP:conf/icse/GengWD00JML24}, we treat the \textit{Others} category, defined as containing vague or non-specific comments, as noisy data and exclude it.
For each remaining category, we insert a corresponding directive into the prompt to guide the agent’s analysis:
\begin{itemize}[leftmargin=8pt]
    \item \textbf{Refactoring}: ``The diff hunk needs to be revised to refactor the code to improve its quality.''
    \item \textbf{Bugfix}: ``The diff hunk needs to be revised to fix one or more bugs.''
    \item \textbf{Testing}: ``The diff hunk needs to be revised since tests for this code must be written.''
    \item \textbf{Logging}: ``The diff hunk needs to be revised to improve the logging of its execution.''
    \item \textbf{Documentation}: ``The diff hunk needs to be revised to be more compliant with the documentation specification.''
\end{itemize}
In the following prompt, the agent is instructed to assess whether the diff hunk requires revision under the "Bugfix" issue category:
\prompt{
\textbf{[\textit{Role}]}~ You are an expert code reviewer. 

\textbf{[\textit{Task}]}~ Give you a diff hunk, your task is to decide whether \textit{\textbf{\textcolor{red}{the diff hunk needs to be revised to fix one or more bugs}}}.
If the response to the above point is True, then write a code review.

\textbf{[\textit{Note}]}~ Note that code changes may be marked: Code lines that begin with a plus (+) sign indicate new code, while those that begin with a minus (-) sign indicate deleted code.
}
Finally, the query code diff hunk is provided as input, prompting the agent to analyze the change from the specified issue-category perspective and generate a corresponding review comment in the expected output format.
\prompt{
For the new diff hunk:

\quad $<Code\ Diff\ Hunk>$

Please output the review comment in the following format:

Review Comment:...
}







\subsection{Discrimination Stage}
\label{sec:discrimination_stage}

In this stage, the identified issues and their associated review comments are thoroughly scrutinized, evaluated, and selected by the critic agent based on factors such as correctness and severity.
The objective of this stage is to simulate the role of an oracle, accurately discerning the correct issue-comment pair and prioritizing it over all other false positives.
The relationship between the category-specific commentator agents and the critic agent is inherently cooperative: the critic’s ability to make precise judgments depends on the quality and diversity of the candidate comments generated by the commentator agents.

In the prompt for the critic agent, we first provide an overview of the framework’s workflow and clearly define its task: to evaluate the candidate review comments based on correctness and severity, and to select the most accurate issue-comment pair as the final output, as shown in the following illustration.
\prompt{
\textbf{[\textit{Role}]}~ You are an expert code reviewer. 

\textbf{[\textit{Task}]}~There is a code review process where a diff hunk is analyzed from five perspectives: 'refactoring', 'bugfix', 'testing', 'logging', and 'documentation', and then the review comments are generated.
After analysis from these different perspectives by category-specific commentator agents, you are required to evaluate the five review comments generated and select the comment that best aligns with the issues present in the target diff hunk.

\textbf{[\textit{Duty}]}~ As a meticulous and harsh critic, your duty is to scrutinize these review comments and evaluate the identified issues with scores in terms of correctness.
}
Next, we provide the critic agent with the query code diff hunk along with the candidate comments generated by the five category-specific commentator agents.
This setup allows the critic to directly compare the insights derived from multiple issue-specific perspectives and assess their relative correctness and severity.
We then prompt the critic agent to analyze the input and select the most appropriate issue category and corresponding review comment, returning the result in the expected output format.
Due to space constraints, the prompt is available in our public repository.

\subsection{Training Stage}

Through the aforementioned two stages, \tech{} enables the generation of review comments by comprehensively analyzing code changes from multiple issue-specific perspectives.
However, due to the inherent complexity of the code review process, it is challenging for LLM-based agents to fully internalize and execute the framework using prompt engineering alone.
Recent studies~\cite{DBLP:journals/corr/abs-2303-14070,huang2023lawyerllamatechnicalreport,DBLP:journals/tosem/YuRSZSWWXW25} have shown that general-purpose LLMs, while equipped with broad linguistic and factual knowledge through large-scale pre-training, require fine-tuning to perform effectively on specialized downstream tasks.
This fine-tuning is essential for aligning model behavior with task-specific requirements and enabling precise reasoning in structured, role-based workflows.
Accordingly, we fine-tune each agent in \tech{} for its designated role, equipping the commentator agents with the ability to generate category-specific review comments and the critic agent with the capability to accurately identify issue categories.

\textbf{Tuning Category-specific Commentator Agents.} 
We begin by partitioning the training dataset, referred to as the Multiple Category Corpus, into five separate Category-specific Corpora based on the issue category of each data instance. 
Each corpus is then used to fine-tune a corresponding commentator agent specialized for that category.
To support this process, we adopt the LlamaFactory framework~\cite{zheng2024llamafactory}, a widely used tool in the open-source community for fine-tuning LLMs.
To minimize computational and storage overhead, we apply low-parameter fine-tuning (i.e., LoRA~\cite{DBLP:journals/corr/abs-2106-09685}), a parameter-efficient fine-tuning method.
LoRA assumes that the parameter updates required during fine-tuning lie in a low-dimensional subspace, allowing the update $\bigtriangleup W$ to be decomposed as:
\begin{equation}
    W' = W_0 + \bigtriangleup W = W_0 + BA
\end{equation}
where, $W'$ represents the fine-tuned parameters, $W_0$ denotes the pre-trained parameters, and $\bigtriangleup W$ signifies the change in the model parameters during fine-tuning. 
Matrices $B \in \mathbb{R}^{d\times r}$ and $A \in \mathbb{R}^{r\times k}$are the low-rank decomposition components, with $d$ and $k$ being the dimensions of the model parameters and $r \ll min(d,k)$. 
During training, only $B$ and $A$ are updated while $W_0$ remains fixed, significantly reducing the number of trainable parameters.
This approach enables efficient fine-tuning of LLMs with minimal resource consumption while preserving performance.

\textbf{Tuning Critic Agent.} 
Unlike the category-specific commentator agents, training the critic agent requires constructing candidate comments from the four non-ground-truth categories for each training instance.
This step is essential to align with the structure expected by the instruction fine-tuning template.
Formally, an instruction tuning template is defined as:
\begin{equation}
    P = \{\mathcal{NL} + x + y + R_x\}
\end{equation}
where $\mathcal{NL}$ is the prompt template described in Section \ref{sec:discrimination_stage}, $x$ and $y$ represent the input code diff hunk and the desired review comment, respectively, and $R_x$ is a set of candidate comments for the other four categories (excluding $c_y$). $R_x$ is defined as:
\begin{equation}
    R_x = \{r_{x,c} | c \in C \backslash \{c_y\}\}
\end{equation}
where $C$ denotes the set of categories, $c_y$ is the category of y and $r_{x,c}$ represents a candidate comment for $x$ in the issue category $c$.
We conjecture that the stronger the relevance between the candidate comment $r_{x,c}$ and the code changes $x$, the more effectively the fine-tuned critic agent can learn the fine-grained differences across various issue categories.
To support this, we introduce a \textit{Candidate Comment Retrieval} approach, which constructs highly relevant candidate comments for each code change. 
This ensures that the critic agent is trained on contextually meaningful and diverse examples.

\textit{Candidate Comment Retrieval.}
Although many studies~\cite{DBLP:journals/corr/abs-2502-11671,DBLP:journals/corr/abs-2408-04125} have explored data augmentation using LLMs, existing research~\cite{DBLP:journals/corr/abs-2408-16502,chan2024balancingcosteffectivenesssynthetic} suggests that its effectiveness diminishes as dataset size increases.
Furthermore, LLM-based augmentation is computationally expensive and time-consuming, making it less practical for large-scale datasets.
To address these limitations, we adopt a simpler and more efficient retrieval-based approach for constructing the training dataset of the critic agent.
Prior work~\cite{DBLP:conf/sigsoft/HongTTA22} has shown that similar code changes often receive similar review comments, even when their surrounding contexts differ.
Inspired by this,  we propose a retrieval-based approach for constructing candidate comments. 
To ensure stronger relevance between the retrieved $r_{x,c}$ and the query diff hunk $x$—for example, ensuring $r_{x,c}$ includes variable names from $x$—we follow prior studies~\cite{DBLP:conf/icse/NashidSM23,DBLP:conf/icse/AhmedPDB24} and employ the BM25 IR algorithm~\cite{DBLP:journals/ftir/RobertsonZ09}. 
Specifically, given a code diff hunk $x$ from the Multiple Category Corpus, \tech{} inputs $x$ into four category-specific commentator agents to obtain the candidate comments $R_x$. Formally, this is defined as:
\begin{equation}
    R_x = \{CCR(x,c) | c \in C \backslash \{c_y\}\}
\end{equation}
where $CCR(x,c)$ denotes the retrieval process, using BM25 to retrieve the most relevant example from the corpora of category $c$ based on the code similarity, and its associated comment is used as the candidate comment $r_{x,c}$.
As the five category-specific corpora are mutually disjoint, there is no risk of candidate comment duplication across categories.
To build a training instance for the critic agent, we combine the instruction prompt $\mathcal{NL}$, the code diff hunk $x$, its ground-truth comment $y$, and the set of generated candidate comments $R_x$.
Finally, we adopt LoRA to fine-tune the critic agent, enhancing its capability on issue-category identification.

\section{Experimental Setup}
\label{sec:exmperiment}

We formulate the following three research questions (RQs):
\begin{itemize}[leftmargin=*]
\item \textbf{RQ1}: How does \tech{} perform compared to the state-of-the-art baselines?
\item \textbf{RQ2}: How much does each component contribute to the overall performance of \tech{}?
  \begin{itemize}
    \item \textbf{RQ2.1}: What is the effect of \textit{Training Stage}?
    \item \textbf{RQ2.2}: What are the effects of the \textit{Generation Stage} and \textit{Discrimination Stage}?
  \end{itemize}
\item \textbf{RQ3}: How does \tech{} perform in human evaluation?
\end{itemize}

\begin{table}[t]

  \centering

  \caption{Overall statistic of Curev datasets\label{tab:dataset}}
  
  \begin{adjustbox}{width=1\linewidth,center}
 \begin{threeparttable}
         \begin{tabular}{l|rcc|rcc}
    \toprule
    \multicolumn{1}{c|}{\multirow{3}[2]{*}{\textbf{Category}}} & \multicolumn{3}{c|}{\textbf{Train}} & \multicolumn{3}{c}{\textbf{Test}} \\
    \cmidrule{2-7} 
          & \multicolumn{1}{c}{\multirow{2}[1]{*}{\textbf{Count}}} & \multicolumn{1}{c}{\textbf{Avg. }} & \multicolumn{1}{c|}{\textbf{Avg. }} & \multicolumn{1}{c}{\multirow{2}[1]{*}{\textbf{Count}}} & \multicolumn{1}{c}{\textbf{Avg. }} & \multicolumn{1}{c}{\textbf{Avg.}} \\
          &       & \multicolumn{1}{c}{\textbf{Code Line}} & \multicolumn{1}{c|}{\textbf{Comment Token}} &       & \multicolumn{1}{c}{\textbf{Code Line}} & \multicolumn{1}{c}{\textbf{Comment Token}} \\
    \midrule
    Refactoring & 10,396 & 20.33  & 22.20  & 3,465  & 19.72  & 22.19  \\
    Bugfix & 2,061  & 18.82  & 25.24  & 687   & 19.74  & 25.42  \\
    Testing & 361   & 31.97  & 23.42  & 120   & 29.85  & 23.30  \\
    Logging  & 146   & 16.16  & 22.01  & 48    & 11.54  & 22.33  \\
    Documentation & 702   & 14.15  & 24.34  & 234   & 16.28  & 23.43  \\
    \midrule
    Total & 13,666 & 20.05  & 22.80  & 4554  & 19.73  & 22.77  \\
    \bottomrule
    \end{tabular}%
\begin{tablenotes}
\item[*]
“Avg.” denotes average values of all instances.
\end{tablenotes}
    \end{threeparttable}
\end{adjustbox}
  \label{tab:dataset}%
\end{table}%

\smallskip
\noindent
\textbf{Dataset.} In this work, we use the multi-category review comment datasets, Curev, introduced by Sghaier et al.~\cite{DBLP:journals/corr/abs-2502-03425} for evaluation. 
Curev is derived from the dataset originally proposed by CodeReviewer~\cite{DBLP:conf/sigsoft/LiLGDJJMGSFS22}, which remains the largest publicly available dataset for code reviews.
The raw dataset contains 176,613 multilingual samples across nine programming languages, including PHP, Ruby, C\#, C, Java, Python, C++, Go, and JavaScript. 
It has been widely used in prior works~\cite{DBLP:conf/icse/TufanoPTPB21,DBLP:conf/icse/TufanoMMPPB22,DBLP:conf/sigsoft/LiLGDJJMGSFS22,DBLP:journals/pacmse/SghaierS24,DBLP:conf/issre/LuYLYZ23,DBLP:journals/tosem/YuRSZSWWXW25,DBLP:journals/tse/NashaatM24}.
Sghaier et al.~\cite{DBLP:journals/corr/abs-2502-03425} used Llama-3.1-70B to evaluate and classify the raw dataset based on the definition of issue category in \cite{DBLP:journals/tse/TufanoDMCB24}.
Then they filtered out the low-quality samples and selected a subset of 20,000 samples to construct the curated datasets.
In this work, following the original configuration, we split the Curev dataset into 75\% for training and 25\% for evaluation, as shown in Table \ref{tab:dataset}, 
ensuring that each sample is a unique code change–comment pair from a single pull request, with no train–test overlap.
Beyond its large scale and multi-language coverage, Curev also exhibits marked class imbalance, with refactoring accounting for 69.3\% of the entire dataset and the remaining categories jointly comprising 30.7\%.
This mirrors real-world issue distributions, where some categories are far more common than others.
As a result, models trained on such data tend to overfit dominant classes (e.g., refactoring) and underperform on rarer but important ones (e.g., testing). 
In this work, we address this methodological challenge directly rather than relying on simple data rebalancing.

\smallskip
\noindent
\textbf{Studied LLMs.}
For practicality (e.g., the high cost of fine-tuning proprietary models like GPT-4o), we excluded proprietary models and evaluated three leading families of open-source LLMs for broader coverage: Qwen2.5-Coder~\cite{qwen2025qwen25technicalreport}, DeepSeek-Coder~\cite{deepseek-coder}, and Llama3~\cite{llama3/url}.
Qwen2.5-Coder is a code-specialized model developed by Alibaba as part of the Qwen 2.5 series. 
It features enhanced performance in programming tasks, achieving state-of-the-art results among open models. 
DeepSeek-Coder is a family of code-focused LLMs developed by DeepSeek, comprising both base and instruction-tuned variants. 
Trained on a diverse dataset including source code and related documentation, it supports long-context inputs up to 16K tokens. 
Llama3 is the latest iteration in the Llama series, developed by Meta. 
It offers enhanced processing power, versatility, and accessibility, with improvements in multilingual support, coding, and reasoning capabilities.

\smallskip
\noindent
\textbf{Baselines.}
Our baseline selection follows established practices in code review research~\cite{DBLP:journals/tosem/YuRSZSWWXW25}, including the most representative and state-of-the-art approaches from relevant categories.
The baselines used for comparison in this study include CodeReviewer~\cite{DBLP:conf/sigsoft/LiLGDJJMGSFS22}, LLaMA-Reviewer~\cite{DBLP:conf/issre/LuYLYZ23},  TufanoLLM~\cite{DBLP:journals/tse/TufanoDMCB24},
and CodeAgent~\cite{DBLP:conf/emnlp/TangKSLLETKB24}:
\begin{itemize}[leftmargin=*,nosep]
    \item \textbf{CodeReviewer} is a pre-trained model to automate code review activities and possesses four specialized tasks to improve its understanding of the review domain.
    To further investigate CodeReviewer’s performance on issue-category identification, we extend it into a variant named \textbf{CodeReviewer-c}, which is retrained to jointly predict both the review category and the corresponding review comment.
    \item \textbf{LLaMA-Reviewer} is an LLM-based approach for automatic code review and uses Parameter-Efficient Fine-Tuning (PEFT) to fine-tune LLaMA for the code review task. 
    In line with CodeReviewer-c, we construct a variant \textbf{LLaMA-Reviewer-c}, which prompts the LLM to generate both the review category and the review comment simultaneously.
    \item \textbf{TufanoLLM} employs a prompt engineering technique that first prompts the LLMs to identify the issue category, followed by generating corresponding review comments.
    \item \textbf{CodeAgent} utilizes a multi-agent framework to detect functionally related defects in pull requests, incorporating inter-agent communication and debate to produce more comprehensive review reports. Although originally operating at the commit level, we adapt it to our function-level setting for a fair comparison.
\end{itemize}
We exclude BitsAI-CR~\cite{DBLP:journals/corr/abs-2501-15134} and CodeMentor~\cite{DBLP:journals/tse/NashaatM24} due to their lack of open-source availability, and 
Carllm~\cite{DBLP:journals/tosem/YuRSZSWWXW25} is omitted as it primarily concerns on fine-tuning LLMs on benchmarks with specific requirements, without proposing generic methodologies on review comment generation.
Additionally, we eliminate CodeDoctor \cite{DBLP:journals/ase/LiWWHWLH25} that generates multi-category review comments at one shot for a given code change, a setting that diverges from the mainstream paradigm of automated code review.
A related line of work is automatic commit-message generation, which summarizes the what and why of code changes, essentially a summarization task. 
In contrast, code review seeks to detect potential issues (e.g., hidden defects), making it a detection task. Due to this task mismatch, we do not treat commit-message generation methods~\cite{DBLP:journals/tse/ZhangQSZZTL24, DBLP:conf/icse/WuWLTYY0L25} as direct baselines.
To ensure a fair comparison and isolate the impact of methodological differences, all LLM-based approaches were implemented using the same underlying LLM.

\smallskip
\noindent
\textbf{Evaluation Metrics.}
We use the following commonly used evaluation metrics to assess the comments generated by \tech{}:
\begin{itemize}[leftmargin=*,nosep]
    \item \textbf{BLEU} (BiLingual Evaluation Understudy)~\cite{DBLP:conf/acl/PapineniRWZ02}  is a precision-based metric computing the weighted geometric mean of modified 1- to 4-gram precisions, with length penalty. 
    In this study, we use BLEU-4, which is widely adopted in code comment generation tasks.

    \item \textbf{ROUGE-L}~\cite{Lin_2004} measures the longest common subsequence (LCS) between the generated and reference comments, focusing on recall and content coverage.
    
    \item \textbf{METEOR}~\cite{DBLP:conf/acl/BanerjeeL05} combines precision and recall using their harmonic mean, offering a more balanced similarity evaluation than BLEU.

    \item \textbf{SBERT}~\cite{DBLP:conf/iwpc/HaqueEBM22} assesses semantic similarity by computing cosine similarity between SentenceBERT~\cite{DBLP:conf/emnlp/ReimersG19} embeddings of generated and reference comments.

    \item \textbf{Pred. Acc.} measures whether \tech{} correctly predicts the issue category of a code diff.
    Formally, let $C_{correct}$ be the number of correct predictions, and $C_{total}$ the total predictions.
    We define prediction accuracy as:
        {
        \begin{equation}
            Pred.~Acc. = \frac{C_{correct}}{C_{total}} \times 100\%
        \end{equation}
        }
\end{itemize}

\smallskip
\noindent
\textbf{Implementation and Environment.}
We implemented our approach using the LlamaFactory~\cite{zheng2024llamafactory} framework. 
All experiments were conducted on a server with four NVIDIA A800 GPUs and 512 GB of memory, running Ubuntu 20.04.2.
Following the prior studies~\cite{DBLP:conf/issre/LuYLYZ23,DBLP:journals/tosem/YuRSZSWWXW25}, 
during training, the model is loaded with float16 precision and trained for 5 epochs with a batch size of 64.
In Low-rank Adaptation (LoRA), we set the learning rate to 0.0003, weight decay to 0.01, LoRA rank to 8, the LoRA scaling factor to 16, LORA dropout to 0.05, and warmup ratio to 0.1.
Following prior work~\cite{DBLP:conf/icse/AhmedPDB24}, we set the temperatures to 0 for well-defined answers from the LLM. 
More details on the specific hyperparameters are available in our materials.



\section{Result}
\label{sec:results}
\subsection{RQ1: Comparison with Baselines}

\begin{table}[t]
\centering
\caption{Performances of \tech{} and baselines on average\label{tab:rq1_average}}
 \begin{adjustbox}{width=.85\linewidth,center}
 \begin{threeparttable}

     \begin{tabular}{l|l|rrrrr}
    \toprule
    \multicolumn{1}{c|}{\textbf{Model}} & \multicolumn{1}{c|}{\textbf{Method}} & \multicolumn{1}{c}{\textbf{BLEU}} & \multicolumn{1}{c}{\textbf{ROUGE-L}} & \multicolumn{1}{c}{\textbf{METEOR}} & \multicolumn{1}{c}{\textbf{SBERT}} & \multicolumn{1}{c}{\textbf{Pred. Acc.}} \\

    \midrule
    \multirow{2}{*}{CodeT5-base} & CodeReviewer & 7.36  & 16.45  & 10.11  & 30.85  & - \\
          & CodeReviewer-c & 6.46  & 14.88  & 7.86  & 19.08  & 76.35\% \\
           \midrule
    \multirow{5}{*}{Llama-3-8B} & LLaMA-Reviewer & 7.64  & 17.70  & 12.32  & 44.42  & - \\
          & LLaMA-Reviewer-c & 7.63  & 17.77  & 11.01  & 40.88  & 51.19\% \\
          & TufanoLLM & 2.83  & 8.56  & 8.37  & 34.52  & 34.89\% \\
          & CodeAgent & 3.80  & 9.68  & 6.50  & 18.49  & - \\
          & \tech{} & \textbf{8.51} & \textbf{19.51} & \textbf{12.73} & \textbf{47.96} & \textbf{56.52\%} \\
    \midrule
    \multirow{5}{*}{Deepseek-Coder-6.7B} & LLaMA-Reviewer & 7.60  & 17.84  & 12.07  & 43.78  & - \\
          & LLaMA-Reviewer-c & 6.22  & 13.93  & 9.68  & 33.71  & 1.84\% \\
          & TufanoLLM & 4.40  & 11.09  & 9.05  & 37.05  & \textbf{72.84\%} \\
          & CodeAgent & 4.39  & 6.95  & 3.78  & 4.68  & - \\
          & \tech{} & \textbf{8.61} & \textbf{19.73} & \textbf{12.97} & \textbf{48.35} & 67.13\% \\
    \midrule
    \multirow{5}{*}{Qwen2.5-Coder-7B} & LLaMA-Reviewer & 7.75  & 17.97  & 12.29  & 44.88  & - \\
          & LLaMA-Reviewer-c & 6.98  & 16.10  & 10.40  & 37.72  & 44.73\% \\
          & TufanoLLM & 2.93  & 11.10  & 11.12  & 42.16  & 41.77\% \\
           & CodeAgent & 5.05  & 8.93  & 3.92  & 7.16  & - \\
          & \tech{} & \textbf{8.69} & \textbf{19.92} & \textbf{13.14} & \textbf{48.58} & \textbf{63.02\%} \\
    \midrule
    \multirow{5}{*}{Qwen2.5-Coder-14B} & LLaMA-Reviewer & 7.91  & 18.30  & 12.55  & 45.50  & - \\
          & LLaMA-Reviewer-c & 8.33  & 19.29  & 11.90  & 44.00  & 52.48\% \\
          & TufanoLLM & 2.70  & 10.24  & 10.67  & 39.91  & 40.21\% \\
          & CodeAgent & 4.16  & 6.16  & 3.93  & 7.38  & - \\
          & \tech{} & \textbf{9.08} & \textbf{20.46} & \textbf{13.50} & \textbf{48.97} & \textbf{54.13\%} \\
    \bottomrule
    
    \end{tabular}%

\end{threeparttable}
\end{adjustbox}
\end{table}

\textbf{Analysis.}
We evaluate \tech{} and four state-of-the-art baselines, TufanoLLM, CodeAgent, CodeReviewer, and LLaMA-Reviewer, as well as two variants (described in Section 4) on the code review comment generation task using four widely adopted metrics: BLEU, ROUGE-L, METEOR, and SBERT.
To ensure model diversity, we select one representative from each major LLM family: LLaMA-3-8B, Deepseek-Coder-6.7B, and Qwen2.5-Coder-7B.
To further explore the impact of model scale on \tech{}’s performance, we also include Qwen2.5-Coder-14B in our experiments.
Due to resource constraints for fine-tuning, we limit our evaluation to models with up to 14B parameters.
To validate the statistical significance of our results, we conducted the Wilcoxon signed-rank test~\cite{woolson2007wilcoxon} at a 0.05 confidence level between \tech{} and the best-performing baseline.
Additionally, to assess the effectiveness of review comment generation across different issue categories, we partition the test dataset into five groups based on issue categories. 
For each category, we report the average prediction accuracy (Pred. Acc.).

\smallskip
\noindent
\textbf{Results.}
Table~\ref{tab:rq1_average} reports the average performance of \tech{} versus the baselines, and Table~\ref{tab:rq1_q} presents the statistical test results.
First, \ul{in terms of textual metrics}, we observe that \tech{} outperforms across all evaluation metrics.
Specifically, \tech{} improves BLEU, ROUGE-L, METEOR, and SBERT by 12.90\%, 10.87\%, 6.32\%, and 8.57\% on average across four different LLMs, respectively, compared to the best-performing baseline, i.e., LLaMA-Reviewer. 
As shown in Table \ref{tab:rq1_q}, statistical testing via the two-sided Wilcoxon signed-rank test confirms significance for all evaluation metrics, with all \textit{p-values} $<$ 7.53e-05.
Furthermore, we find that \tech{} slightly outperforms LLaMA-Reviewer in METEOR across these four LLMs. 
This difference may be attributed to the nature of the METEOR metric, which is sensitive to synonyms and stem variations but also penalizes non-contiguous word order. 
Consequently, \tech{} may generate more flexible or restructured phrasing, which makes its scores slightly lower on METEOR.
This is further supported by the SBERT results, where \tech{} consistently achieves the highest scores among all baselines, indicating its superior ability to generate semantically accurate review comments.
In contrast, TufanoLLM shows weak performance. 
This may be due to its lack of access to sufficient in-context examples, which makes it difficult to accurately identify the issue categories in the code diffs, resulting in divergent and imprecise review comments.
Similarly, CodeAgent achieves poor performance due to its focus on security analysis, making it hard to detect other categories of issues in the code diffs.
Moreover, we observe that CodeReviewer-c and LLaMA-Reviewer-c, which incorporate issue-category predictions, generally exhibit a slight performance decline compared to their original counterparts.
This degradation likely results from the increased complexity of combining classification with generation, which transforms the task from a single-objective generation problem into a more challenging classification-plus-generation setting.

\begin{table}[t]
\centering
\caption{Wilcoxon Signed-Rank Test P-Values for Metric Comparisons Between \tech{} and LLaMA-Reviewer
\label{tab:rq1_q}}
 \begin{adjustbox}{width=.7\linewidth,center}
 \begin{threeparttable}

    \begin{tabular}{lllll}
    \toprule
    \multicolumn{1}{c}{\textbf{LLM}} & \multicolumn{1}{c}{\textbf{BLEU}} & \multicolumn{1}{c}{\textbf{ROUGE-L}} & \multicolumn{1}{c}{\textbf{METEOR}} & \multicolumn{1}{c}{\textbf{SBERT}} \\
    \midrule
Llama-3-8B & $6.86\times10^{-35}$ & $1.44\times10^{-29}$ & $7.53\times10^{-5}$ & $1.40\times10^{-27}$ \\
Deepseek-Coder-6.7B & $2.05\times10^{-33}$ & $5.07\times10^{-28}$ & $5.71\times10^{-15}$ & $4.00\times10^{-43}$ \\
Qwen2.5-Coder-7B & $1.46\times10^{-27}$ & $8.04\times10^{-30}$ & $4.78\times10^{-12}$ & $4.09\times10^{-28}$ \\
Qwen2.5-Coder-14B & $8.07\times10^{-39}$ & $7.84\times10^{-37}$ & $4.98\times10^{-16}$ & $2.50\times10^{-26}$ \\

    \bottomrule
    \end{tabular}%

\end{threeparttable}
\end{adjustbox}
\end{table}

\ul{In terms of the model diversity and scale}, when comparing models of similar sizes from different families (6.7B, 7B, and 8B), we observe that the general-purpose model LLaMA-3, which is not specialized for the code task, performs the worst, while Qwen2.5-Coder and DeepSeek-Coder show complementary strengths, each excelling in different aspects.
These findings suggest that when model sizes are comparable, performance under \tech{} is potentially influenced more by model specialization. 
Conversely, when comparing models of different sizes within the same family (i.e., Qwen2.5-Coder 7B vs. 14B), the larger 14B variant outperforms the 7B model by 0.39, 0.54, 0.36, and 0.39 points on BLEU, ROUGE-L, METEOR, and SBERT, respectively. 
This indicates that as model size increases, \tech{} likely further amplifies its performance.

\begin{table}[t]
\centering
\caption{Performances of \tech{} and baselines on each category on Deepseek-Coder-6.7B\label{tab:rq1_category}}

 \begin{adjustbox}{width=0.8\linewidth,center}
 \begin{threeparttable}
    \begin{tabular}{l|l|rrrrr}
    \toprule
    \multicolumn{1}{c|}{\textbf{Category}} & \multicolumn{1}{c|}{\textbf{Method}} & \multicolumn{1}{c}{\textbf{BLEU}} & \multicolumn{1}{c}{\textbf{ROUGE-L}} & \multicolumn{1}{c}{\textbf{METEOR}} & \multicolumn{1}{c}{\textbf{SBERT}} & \multicolumn{1}{c}{\textbf{Pred. Acc.}} \\
    \midrule
    \multirow{7}{*}{Refactoring} & CodeReviewer & 7.65  & 17.02  & 10.85  & 35.42  & - \\
          & CodeReviewer-c & 6.70  & 15.47  & 8.26  & 20.00  & \textbf{97.72\%} \\
          & LLaMA-Reviewer & 7.58  & 17.84  & 12.03  & 43.44  & - \\
          & LLaMA-Reviewer-c & 6.31  & 14.00  & 9.78  & 33.60  & 2.11\% \\
          & TufanoLLM & 4.36  & 10.87  & 8.97  & 36.70  & 95.53\% \\
           & CodeAgent & 4.35 & 6.95 & 3.88 & 3.02  & - \\
          & \tech{} & \textbf{8.67} & \textbf{19.98} & \textbf{13.08} & \textbf{48.23} & 82.48\% \\
    \midrule
    \multirow{7}{*}{Bugfix} & CodeReviewer & 6.03  & 13.29  & 7.15  & 15.35  & - \\
          & CodeReviewer-c & 5.33  & 11.74  & 6.22  & 15.59  & 4.95\% \\
          & LLaMA-Reviewer & 7.49  & 17.13  & 11.91  & 44.83  & - \\
          & LLaMA-Reviewer-c & 5.59  & 12.87  & 8.91  & 32.99  & 0.58\% \\
          & TufanoLLM & 4.54  & 12.00  & 9.34  & 38.91  & 0.87\% \\
           & CodeAgent & 4.31 & 6.87 & 3.78 & 5.20  & - \\
          & \tech{} & \textbf{8.14} & \textbf{17.66} & \textbf{12.28} & \textbf{49.29} & \textbf{21.69\%} \\
    \midrule
    \multirow{7}{*}{Testing} & CodeReviewer & 7.43  & 16.17  & 8.56  & 19.49  & - \\
          & CodeReviewer-c & 5.75  & 11.70  & 6.14  & 18.52  & 0.00\% \\
          & LLaMA-Reviewer & 7.60  & 17.13  & 11.91  & 41.17  & - \\
          & LLaMA-Reviewer-c & 5.77  & 12.98  & 8.38  & 28.30  & 0.83\% \\
          & TufanoLLM & 4.60  & 10.98  & 8.39  & 30.87  & 0.00\% \\
          & CodeAgent & 4.45 & 6.38 & 3.49 & 6.53  & - \\
          & \tech{} & \textbf{8.15} & \textbf{19.44} & \textbf{11.99} & \textbf{43.73} & \textbf{14.17\%} \\
    \midrule
    \multirow{7}{*}{Logging} & CodeReviewer & 6.66  & 15.96  & 9.42  & 31.91  & - \\
          & CodeReviewer-c & 5.86  & 13.84  & 7.54  & 27.93  & 0.00\% \\
          & LLaMA-Reviewer & 7.55  & 19.56  & \textbf{12.54} & \textbf{49.94} & - \\
          & LLaMA-Reviewer-c & 6.09  & 15.06  & 10.26  & 42.74  & 0.02\% \\
          & TufanoLLM & 4.39  & 11.67  & 9.61  & 41.91  & 2.08\% \\
           & CodeAgent & 4.44 & 6.41 & 3.30 & 6.68  & - \\
          & \tech{} & \textbf{8.26} & \textbf{19.95} & 12.52  & 49.37  & \textbf{8.33\%} \\
    \midrule
    \multirow{7}{*}{Documentation} & CodeReviewer & 7.12  & 17.59  & 8.81  & 14.37  & - \\
          & CodeReviewer-c & 6.62  & 17.16  & 7.71  & 13.97  & \textbf{24.36\%} \\
          & LLaMA-Reviewer & 8.20  & 19.96  & 13.43  & 45.83  & - \\
          & LLaMA-Reviewer-c & 7.11  & 16.25  & 11.05  & 38.48  & 0.02\% \\
          & TufanoLLM & 4.63  & 11.67  & 9.56  & 38.94  & 0.00\% \\
           & CodeAgent & 4.77 & 7.52 & 3.53 & 9.80  & - \\
          & \tech{} & \textbf{9.36} & \textbf{22.22} & \textbf{13.99} & \textbf{49.61} & 12.39\% \\
    \midrule
    \multirow{7}{*}{Average} & CodeReviewer & 7.36  & 16.45  & 10.11  & 30.85  & - \\
          & CodeReviewer-c & 6.46  & 14.88  & 7.86  & 19.08  & \textbf{76.35\%} \\
          & LLaMA-Reviewer & 7.60  & 17.84  & 12.07  & 43.78  & - \\
          & LLaMA-Reviewer-c & 6.22  & 13.93  & 9.68  & 33.71  & 1.84\% \\
          & TufanoLLM & 4.40  & 11.09  & 9.05  & 37.05  & 72.84\% \\
           & CodeAgent & 4.39  & 6.95  & 3.78  & 4.68  & - \\
          & \tech{} & \textbf{8.61} & \textbf{19.73} & \textbf{12.97} & \textbf{48.35} & 67.13\% \\
    \bottomrule
   
    \end{tabular}%

\end{threeparttable}
\end{adjustbox}
\end{table}

We further analyzed the average performance across the five issue categories, as shown in Table~\ref{tab:rq1_category}.
Given the importance of accurate issue-category identification for fine-grained analysis and downstream code refinement, we selected Deepseek-Coder-6.7B for detailed examination.
This model achieved the highest prediction accuracy (Pred. Acc.), demonstrating a strong balance between review comment quality and classification performance.
In terms of per-category performance, \tech{} outperforms all baselines across five metrics in the majority of issue categories.
However, it underperforms LLaMA-Reviewer on METEOR and SBERT within the Logging category, likely due to the underrepresentation of Logging samples in the critic agent’s training set, which affects its ability to accurately identify logging-related issues.
A similar trend appears in the Refactoring category, where \tech{} falls behind CodeReviewer-c and TufanoLLM in issue-category prediction accuracy.
Further analysis shows that both models exhibit a strong bias toward labeling most changes as Refactoring.
Since Refactoring is the most frequent category in the test set, this bias artificially inflates their prediction accuracy.
This behavior is rooted in the dominance of Refactoring scenarios in real-world datasets, which biases the LLMs during pre-training.
In contrast, \tech{} mitigates this bias through targeted fine-tuning.
By leveraging category-specific features from both code and comments, it achieves more balanced and fair classification across all issue categories.
Notably, \tech{} demonstrates superior performance on challenging issue types, achieving 21.69\% accuracy in the Bugfix category, where existing approaches typically achieve less than 5\%.
By contrast, CodeAgent performs the worst across all categories. 
Despite its multi-agent architecture, it lacks category-specific training and targeted knowledge alignment.
As a result, agents likely fail to capture issue-dependent characteristics, yielding weak overall performance.

\finding{
    \textbf{Answering RQ1:}~ 
    Compared to the best-performing baselines, \tech{} significantly improves the average performance of BLEU, ROUGE-L, METEOR, and SBERT by 12.90\%, 10.87\%, 6.32\%, and 8.57\% across four LLMs.
    Moreover, \tech{} demonstrates superior performance on tackling challenging issue types (i.e., bug-fix, testing, and logging). 
    }

\subsection{RQ2: Ablation Experiment}

This section evaluates the contribution of each component in the \textit{Generation}, \textit{Discrimination}, and \textit{Training} stages to the overall performance of \tech{}.
Since the \textit{Generation} and \textit{Discrimination Stages} focus on producing review comments, while the \textit{Training Stage} equips them with the necessary capabilities, we first analyze the \textit{Training Stage} in RQ2.1, followed by the analysis of the \textit{Generation} and \textit{Discrimination Stages} in RQ2.2.

\smallskip

\noindent
\textbf{RQ2.1: Effect of Training Stage}
\label{sec:ablation_trainng}
\begin{table}[t]
\centering
\caption{Performances of \tech{} and variants on average\label{tab:rq2_average}}
 \begin{adjustbox}{width=.8\linewidth,center}
 \begin{threeparttable}
    \begin{tabular}{l|l|rrrrr}
    \toprule
    \multicolumn{1}{c|}{\textbf{Model}} & \multicolumn{1}{c|}{\textbf{Method}} & \multicolumn{1}{c}{\textbf{BLEU}} & \multicolumn{1}{c}{\textbf{ROUGH-L}} & \multicolumn{1}{c}{\textbf{METEOR}} & \multicolumn{1}{c}{\textbf{SBERT}} & \multicolumn{1}{c}{\textbf{Pred. Acc.}} \\
    \midrule
    \multirow{5}{*}{Llama-3-8B} & \tech{} & \textbf{8.51} & \textbf{19.51} & \textbf{12.73} & \textbf{47.96} & \textbf{56.52\%} \\
          & \textit{w/o} SFT & 4.43  & 9.70  & 5.79  & 20.90  & 42.89\% \\
          & \textit{w/o} SFT-MCCA & 6.32  & 13.54  & 8.31  & 34.23  & 31.42\% \\
          & \textit{w/o} SFT-CA & 4.97  & 11.29  & 7.14  & 26.10  & 44.69\% \\
          & \textit{w/o} CCR & 8.41  & 19.10  & 11.83  & 44.95  & 47.85\% \\
    \midrule
    \multirow{5}{*}{Deepseek-Coder-6.7B} & \tech{} & \textbf{8.61} & \textbf{19.73} & \textbf{12.97} & \textbf{48.35} & \textbf{67.13\%} \\
          & \textit{w/o} SFT & 5.09  & 11.24  & 6.36  & 16.82  & 60.14\% \\
          & \textit{w/o} SFT-MCCA & 5.84  & 13.08  & 7.08  & 23.59  & 49.60\% \\
          & \textit{w/o} SFT-CA & 7.03  & 15.97  & 10.23  & 38.47  & 63.57\% \\
          & \textit{w/o} CCR & 8.36  & 19.21  & 12.32  & 46.30  & 65.48\% \\
    \midrule
    \multirow{5}{*}{Qwen2.5-Coder-7B} & \tech{} & \textbf{8.69} & \textbf{19.92} & \textbf{13.14} & \textbf{48.58} & \textbf{63.02\%} \\
          & \textit{w/o} SFT & 7.07  & 15.45  & 10.53  & 41.63  & 52.83\% \\
          & \textit{w/o} SFT-MCCA & 7.49  & 16.92  & 10.96  & 43.69  & 56.08\% \\
          & \textit{w/o} SFT-CA & 8.27  & 18.28  & 12.12  & 45.68  & 47.43\% \\
          & \textit{w/o} CCR & 8.62  & 19.54  & 12.59  & 46.48  & 54.61\% \\
    \midrule
    \multirow{5}{*}{Qwen2.5-Coder-14B} & \tech{} & \textbf{9.08} & \textbf{20.46} & \textbf{13.50} & \textbf{48.97} & 54.13\% \\
          & \textit{w/o} SFT & 7.67  & 17.61  & 10.78  & 42.24  & 58.21\% \\
          & \textit{w/o} SFT-MCCA & 7.76  & 17.73  & 11.43  & 45.65  & 38.45\% \\
          & \textit{w/o} SFT-CA & 8.94  & 20.18  & 12.92  & 46.84  & \textbf{64.16\%} \\
          & \textit{w/o} CCR & 8.97  & 20.11  & 12.76  & 46.36  & 61.35\% \\
    \bottomrule
    \end{tabular}%

\end{threeparttable}
\end{adjustbox}
\end{table}

\smallskip

\noindent
\textbf{Analysis.}
To assess the contributions of each core component of \textit{Training Stage} in \tech{}, we conducted an ablation study comparing \tech{} with four variants:
\begin{itemize}[leftmargin=10pt,nosep]
    \item \textit{w/o} SFT: 
    Replacing Supervised Fine-Tuning (SFT) with few-shot learning for both the category-specific commentator agents and the critic agent, aiming to assess the necessity of fine-tuning.
    \item \textit{w/o} SFT-MCCA: 
    Replacing the Supervised Fine-Tuned Multiple Category-specific Commentator Agents (SFT-MCCA) with their few-shot learning counterparts, isolating the impact of fine-tuning specifically for the commentator agents.
    \item  \textit{w/o} SFT-CA: Replacing the Supervised Fine-Tuned Critic Agent (SFT-CA) with a few-shot learning version, to assess the importance of fine-tuning specifically for the critic agent.
    \item \textit{w/o} CCR: 
    Removing the Candidate Comment Retrieval (CCR) during the construction of the critic agent’s training dataset. Instead, it uses candidate comments generated by the fine-tuned category-specific commentator agents to evaluate the role of CCR in improving issue-category identification.
\end{itemize}
Following prior works~\cite{DBLP:conf/kbse/AhmedD22,DBLP:conf/iclr/PatelLRCRC23,DBLP:conf/icse/AhmedPDB24}, our experiments adopt a 3-shot setting and utilize BM25 similarity~\cite{DBLP:journals/ftir/RobertsonZ09} to retrieve relevant code changes for demonstration purposes.

\smallskip
\noindent
\textbf{Results.}
Table \ref{tab:rq2_average} presents the related results.
Overall, all variants exhibit varying degrees of performance degradation compared to \tech{}.
Specifically, the performance degradation of \tech{} \textit{w/o} SFT highlights the necessity of fine-tuning both the category-specific commentator agents and the critic agent. 
This step is critical for enhancing these agents to generate accurate category-specific review comments and to perform effective issue-category identification. 
At the same time, the drop in performance observed in \tech{} \textit{w/o} SFT-MCCA and \tech{} \textit{w/o} SFT-CA suggests that while those agents can acquire a basic level of task capability through few-shot learning, they still underperform compared to the fine-tuned versions.
Interestingly, on Qwen2.5-Coder-14B, \tech{} \textit{w/o} SFT-CA achieves a higher prediction accuracy and close performance comparable to \tech{} across the other three metrics.
This can be attributed to Qwen2.5-Coder-14B’s strong code understanding capabilities, which enable it to achieve good performance even with limited supervision.
Nevertheless, fine-tuning remains essential for equipping smaller models with capabilities that exceed those of much larger models.

Furthermore, we observe that \tech{} \textit{w/o} CCR outperforms \tech{} in prediction accuracy on Qwen2.5-Coder-14B. 
This may be because, compared to smaller LLMs, the category-specific commentator agents based on Qwen2.5-Coder-14B generate higher-quality candidate comments used to train the critic agent with stylistic consistency closer to the ground truth. 
Such consistency enables the critic agent to better focus on semantic distinctions—such as reasoning about variables and logic structures—rather than being distracted by stylistic variations. 
As a result, even without CCR, the critic agent trained on Qwen2.5-Coder-14B demonstrates strong issue-category identification capabilities. 
In contrast, with smaller LLMs, the quality of generated candidate comments is generally lower than that of retrieved comments, leading to poorer performance without CCR.
Therefore, \tech{} \textit{w/o} CCR tends to perform better on more capable models, such as Qwen2.5-Coder-14B.

\finding{
    \textbf{Answering RQ2.1:}~ 
    Both Supervised Fine-Tuning and Comment Candidate Retrieval significantly enhance the \textit{Training Stage} of \tech{}, 
    enabling agents to achieve more accurate issue-category identification and generate higher-quality review comments.
    }
\smallskip

\noindent
\textbf{RQ2.2: Effect of Generation and Discrimination Stages}
\label{sec:ablation_Generation_Discrimination}

\begin{table}[t]
\centering
\caption{Performances of different configurations of two
types of agent in \tech{} on average\label{tab:rq3_all}}
 \begin{adjustbox}{width=.7\linewidth,center}
 \begin{threeparttable}
  \begin{tabular}{l|l|rrrrr}
    \toprule
    \multicolumn{1}{c|}{\textbf{Model}} & \multicolumn{1}{c|}{\textbf{Method}} & \multicolumn{1}{c}{\textbf{BLEU}} & \multicolumn{1}{c}{\textbf{ROUGE-L}} & \multicolumn{1}{c}{\textbf{METEOR}} & \multicolumn{1}{c}{\textbf{SBERT}} & \multicolumn{1}{c}{\textbf{Pred. Acc.}} \\
    \midrule
    \multirow{3}[2]{*}{Llama-3-8B} & \tech{} & \textbf{8.51} & \textbf{19.51} & 12.73  & \textbf{47.96} & \textbf{56.52\%} \\
          & \textit{w/} SFA & 7.82  & 18.03  & \textbf{12.97} & 47.94  & 42.64\% \\
          & \textit{w/} MSC & 3.83  & 13.36  & 12.51  & 43.92  & - \\
    \midrule
    \multirow{3}[2]{*}{Deepseek-Coder-6.7B} & \tech{} & \textbf{8.61} & \textbf{19.73} & 12.97  & \textbf{48.35} & \textbf{67.13\%} \\
          & \textit{w/} SFA & 8.27  & 18.94  & \textbf{12.99} & 47.93  & 60.32\% \\
          & \textit{w/} MSC & 8.13  & 18.35  & 11.62  & 44.39  & - \\
    \midrule
    \multirow{3}[2]{*}{Qwen2.5-Coder-7B} & \tech{} & \textbf{8.69} & \textbf{19.92} & 13.14  & 48.58  & \textbf{63.02\%} \\
          & \textit{w/} SFA & 8.34  & 19.01  & 13.17  & 48.52  & 59.95\% \\
          & \textit{w/} MSC & 4.38  & 16.02  & \textbf{15.63} & \textbf{50.02} & - \\
    \midrule
    \multirow{3}[2]{*}{Qwen2.5-Coder-14B} & \tech{} & \textbf{9.08} & \textbf{20.46} & 13.50  & 48.97  & \textbf{54.13\%} \\
          & \textit{w/} SFA & 8.05  & 18.65  & 13.37  & 48.49  & 47.83\% \\
          & \textit{w/} MSC & 4.47  & 16.27  & \textbf{15.40} & \textbf{49.73} & - \\
    \bottomrule
    \end{tabular}%

\end{threeparttable}
\end{adjustbox}
\end{table}

\smallskip
\noindent
\textbf{Analysis.}
To further evaluate the effectiveness of the current configuration of the category-specific commentator agents in the \textit{Generation Stage} and the critic agent in the \textit{Discrimination Stage}, we designed alternative configurations as comparative variants:
\begin{itemize}[leftmargin=10pt,nosep]
\item \textit{with} SFA: 
    This variant replaces the Multiple Category-specific Commentator Agents (MCCA) with a Single Fusion Agent (SFA), which is fine-tuned on the entire multi-category corpus and is responsible for generating review comments across all issue categories.
    This setup allows us to assess the importance of training separate agents for each specific issue category.
\item \textit{with} MSC: 
    Instead of selecting a single best review comment from the candidate comments, this variant allows the critic to select Multiple Suitable Comments (MSC) and merge them.
    Specifically, the critic agent's prompt is modified as "Please read these review comments and merge the suitable review comments."
    This variant enables us to investigate whether merging multiple review comments can lead to better performance.
\end{itemize}

\smallskip
\noindent
\textbf{Results.}
The results under different configurations of the two types of agents in \tech{} are shown in Table~\ref{tab:rq3_all}.
Overall, \tech{} outperforms both \tech{} \textit{with} SFA and \tech{} \textit{with} MSC across the majority of evaluation metrics.
Interestingly, \tech{} \textit{with} SFA shows slightly higher METEOR scores.
This is primarily due to the fusion agent, trained on the entire dataset, tending to generate comments with more common tokens.
Consequently, as the critic agent misclassifies the issue category, the generated comments may still exhibit greater token-level overlap with the ground truth, leading to inflated METEOR scores despite being semantically inaccurate.
Similarly, we observe that \tech{} \textit{with} MSC consistently yields lower BLEU and ROUGE-L scores across all four LLMs. 
However, for both versions of Qwen2.5-Coder, this variant achieves higher scores on METEOR and SBERT.
This discrepancy can be attributed to the merging of multiple review comments, which decreases token-level similarity with the ground truth, leading to lower BLEU and ROUGE-L scores, while enhancing semantic richness, thereby improving performance on metrics that emphasize semantic similarity, such as METEOR and SBERT.
These findings indicate that both the commentator agents and the critic agent are essential to \tech{}’s effectiveness.
Notably, when comparing the performance differences across the two variants, \tech{} \textit{with} MSC exhibits more pronounced changes, showing significant declines in BLEU and ROUGE-L, alongside noticeable improvements in METEOR and SBERT.
This suggests that the critic agent plays a particularly critical role in the overall system.

\finding{
    \textbf{Answering RQ2.2:}~
   The category-specific commentator agents and the critic agent both contribute to \tech{}’s effectiveness in terms of textual similarity and prediction accuracy, with the critic agent playing the most crucial role.
}

\subsection{RQ3: Human Evaluation}

\smallskip
\noindent
\textbf{Analysis.}
Although the four existing evaluation metrics measure lexical and semantic differences between generated comments and ground truth, they often fail to capture real semantic disparities.
To comprehensively evaluate the quality of review comments generated by various approaches, we rigorously conducted a human evaluation.
Following prior works~\cite{DBLP:conf/icse/MuCSWW23,DBLP:journals/corr/abs-2502-03425,DBLP:journals/tosem/YuRSZSWWXW25}, we randomly selected 384 samples (achieving a 95\% confidence level with a confidence interval of less than 3\%) while maintaining proportional representation across issue categories in the test set (ensuring at least one sample from each category). 
This process resulted in 1,920 generated comments for evaluation from five approaches (i.e., CodeReviewer, LLaMA-Reviewer, TufanoLLM, CodeAgent, and \tech{}) on Deepseek-Coder-6.7B.
Consistent with prior studies~\cite{DBLP:journals/corr/abs-2502-03425}, we adopted a reference-free evaluation, which allows evaluators to exercise subjective judgment, recognize reasonable comments beyond reference matches, assess additional dimensions (e.g., readability), and reduce bias from poor-quality references~\cite{DBLP:conf/eval4nlp/AkkasiFK23,DBLP:conf/emnlp/GigantGDD24}.
We recruited two participants with over five years of Java development experience, who are not co-authors of this paper. 
Among the 384 randomly selected samples, the first evaluator reviewed the first 242 samples, and the second evaluator reviewed the last 242, with 100 overlapping samples in the middle used to measure inter-rater agreement via Cohen's kappa.
We also ensured that each evaluator assessed an equal number of cases from each issue category to provide balanced evaluation coverage.
To ensure fairness, evaluators were not informed of the source of the comments. 
Each comment was rated across three dimensions:
(1) \textbf{Readability}, which reflects  the fluency, clarity, specificity, and ease of understanding of generated review comments;
(2) \textbf{Accuracy}, which reflects whether they clearly point out code issues rather than being vague, and whether there are any incorrect suggestions;
(3) \textbf{Category-Matching}, which reflects whether the review comments are relevant to the target issue categories.
All ratings were given on a 5-point Likert scale: 1 for poor, 2 for marginal, 3 for acceptable, 4 for good, and 5 for excellent.
The Cohen's Kappa coefficient between the two evaluators is 0.74, indicating substantial agreement and confirming the reliability of their assessments.

\begin{table}[t]
\centering
\caption{Results of the human evaluation on 1,920 generated comments from \tech{} and four baselines\label{tab:rq4}}
 \begin{adjustbox}{width=0.5\linewidth,center}
 \begin{threeparttable}

    \begin{tabular}{llcc}
    \toprule
    \makecell[c]{\textbf{Metric}} & \makecell[l]{\textbf{Method}} & \multicolumn{1}{c}{\textbf{Avg.}} & \multicolumn{1}{c}{\textbf{Std.}} \\
    \midrule
    \multirow{5}{*}{Readability} & CodeReviewer  & 2.3   & 1.2  \\
          & LLaMA-Reviewer & 3.1   & 1.0  \\
          & TufanoLLM & 2.6   & 1.3  \\
          & CodeAgent & 1.3 & 0.8 \\
          & \tech{} & 3.8   & 0.7  \\
    \midrule
    \multirow{5}{*}{Accuracy} & CodeReviewer  & 2.3   & 1.3  \\
          & LLaMA-Reviewer & 2.8   & 1.3  \\
          & TufanoLLM & 2.3   & 1.3  \\
          & CodeAgent & 1.3 & 0.8 \\
          & \tech{} & 3.6   & 1.0  \\
    \midrule
    \multirow{5}{*}{Category-Matching} & CodeReviewer  & 2.2   & 1.1  \\
          & LLaMA-Reviewer & 2.8   & 0.8  \\
          & TufanoLLM & 2.5   & 1.3  \\
          & CodeAgent &  1.2 & 0.6 \\
          & \tech{} & 3.5   & 0.9  \\
    \bottomrule
    \end{tabular}%
\end{threeparttable}
\end{adjustbox}
\end{table}

\begin{figure}[t]
    \includegraphics[width=1\columnwidth]{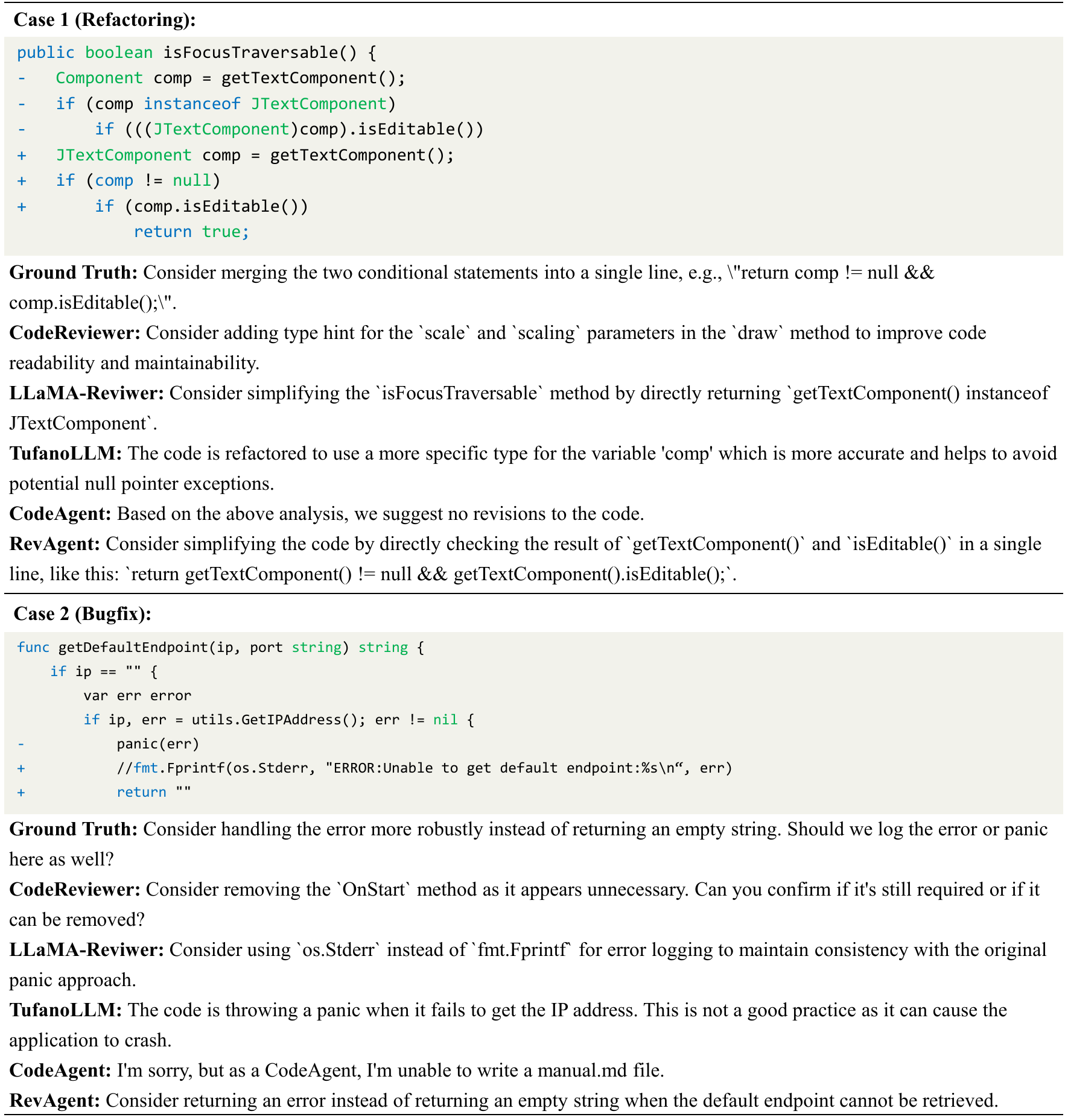}
	\centering
    \caption{Generated comments by \tech{} and baselines} 
    \label{fig:case}
\end{figure}

\smallskip
\noindent
\textbf{Results.}
Table~\ref{tab:rq4} summarizes the related human evaluation results.
Overall, \tech{} outperforms all baselines across the three evaluation dimensions: Readability, Accuracy, and Category-Matching.
For example, \tech{} outperforms LLaMA-Reviewer by 0.7, 0.8, and 0.7 points, respectively.
This demonstrates that \tech{} is capable of generating review comments that align well with human judgment.
In particular, \tech{} significantly surpasses the baselines in both generating accurate review comments and correctly identifying the issue categories present in code diffs.
Moreover, its lower standard deviation in Readability and Accuracy indicates more consistent performance across different cases.
We also observe that TufanoLLM, due to its open-ended generation style, achieves better Readability and Category-Matching scores than CodeReviewer.
In contrast, LLaMA-Reviewer, which benefits from supervised fine-tuning, consistently outperforms the CodeT5-based CodeReviewer across all three evaluation metrics.
Finally, CodeAgent performs the worst among all methods, primarily because its excessive openness, where all steps are autonomously executed by agents without explicit guidance, leads to unstable and inaccurate results when handling complex code review tasks.
Figure~\ref{fig:case} presents a refactoring case and a bugfix case.
In the refactoring case, the code diff contains multiple nested conditional statements that should be merged.
While the other baselines failed to identify the core issue, focusing instead on aspects such as variable declarations and function signatures, \tech{} correctly detects the problem of deeply nested conditions and provides a feasible refactoring suggestion.
In the bug-fix case, \tech{} successfully identifies the issue with the abnormal return value in the code diff, further highlighting its superiority in precise issue detection.
\finding{
    \textbf{Answering RQ3:}~
    In human evaluation, compared to the baselines, \tech{} achieves the highest scores on Readability, Accuracy, and Category-Matching, demonstrating its effectiveness in generating high-quality comments that help developers identify and understand code issues.
}

\section{Discussion}
\label{sec:discussion}

\textbf{Inference Consumption Analysis.}
Although the preceding RQ results demonstrate the effectiveness of our approach, the efficiency of the agent-based framework, particularly for LLM-based agents, should also be considered for practical applicability.
To assess the efficiency of \tech{}, we measured both average token consumption and inference time across four LLMs.
As shown in Table~\ref{tab:discussion_all}, \tech{} increases token usage by $8.68\times$ and $4.47\times$ compared to LLaMA-Reviewer and TufanoLLM, respectively, but only $0.06\times$ that of CodeAgent.
While the average token count is relatively high (2,468 tokens on average), it remains acceptable in practice, as we utilize open-source LLMs that do not incur additional token-based costs.
Moreover, \tech{}’s token consumption is significantly lower than that of the multi-agent-based CodeAgent.
In terms of inference time, \tech{} introduces a $1.97\times$ latency increase compared to LLaMA-Reviewer, but operates $0.46\times$  faster than TufanoLLM.
Despite employing an agent-based architecture, the additional 0.038 seconds per prediction compared to LLaMA-Reviewer falls within a reasonable trade-off.
Furthermore, compared with CodeAgent’s multi-agent design, our framework is considerably lighter, achieving substantially lower inference time due to its streamlined and well-coordinated agent interactions.
In summary, considering the substantial performance improvements achieved by \tech{}, the framework maintains a favorable balance between effectiveness and efficiency, making it suitable for practical large-scale code review applications.

\begin{table*}[t]

  \centering

  \caption{The average inference consumption of \tech{} and baselines.\label{tab:discussion_all}}
  
  \begin{adjustbox}{width=1\linewidth,center}
 \begin{threeparttable}
    \begin{tabular}{l|cccc|cccc}
    \toprule
    \multicolumn{1}{c|}{\multirow{2}[4]{*}{\textbf{LLM}}} & \multicolumn{4}{c|}{\textbf{Time Consumption}} & \multicolumn{4}{c}{\textbf{Token Consumption}} \\
\cmidrule{2-9}          & \multicolumn{1}{c}{LLaMA-Reviewer} & \multicolumn{1}{c}{TufanoLLM} & \multicolumn{1}{c}{CodeAgent} & \multicolumn{1}{c|}{\tech{}} & \multicolumn{1}{c}{LLaMA-Reviewer} & \multicolumn{1}{c}{TufanoLLM} & \multicolumn{1}{c}{CodeAgent} & \multicolumn{1}{c}{\tech{}} \\
    \midrule
    LLaMA-3-8B & 0.016  & 0.055  & 212.603 & 0.049  & 232   & 420   & 39,143 & 2,338  \\
    Deepseek-Coder-6.7B & 0.019  & 0.202 & 238.670 & 0.056  & 294   & 532 & 99,713  & 2,796  \\
    Qwen2.5-Coder-7B & 0.015  & 0.027 & 183.868 & 0.044  & 247   & 427 & 6,887  & 2,371  \\
    Qwen2.5-Coder-14B & 0.026  & 0.047 & 163.263 & 0.077  & 247   & 427 & 7,096  & 2,366  \\
    \midrule
    Average & 0.019  & 0.083 & 240.417 & 0.057  & 255   & 452 & 39,984  & 2,468  \\
    \bottomrule
    \end{tabular}%
    \end{threeparttable}
\end{adjustbox}
\end{table*}%

\smallskip
\noindent
\textbf{Role of Issue Prediction in Review Comment Generation.}
To further investigate the influence of issue-category prediction on comment generation, we analyzed the quality of review comments generated by \tech{} under both correct and incorrect issue predictions using Deepseek-Coder-6.7B.
Specifically, when the issue category is correctly predicted, \tech{} achieves BLEU, ROUGE-L, METEOR, and SBERT scores of 9.08, 20.68, 13.50, and 49.25, respectively.
In cases of misprediction, these scores decrease to 7.65, 17.79, 11.90, and 46.51, indicating a consistent decline in performance across all metrics.
This degradation can be attributed to the role of issue categories in choosing the correct commentator agents to focus on the most relevant issues of the code change. 
Accurate prediction ensures that the corresponding category-specific agent is better aligned with the underlying issue in the code change, thereby facilitating the generation of precise and contextually appropriate comments.
Conversely, incorrect predictions may lead to an ill-suited agent, resulting in comments that are either overly generic or fail to address the core issue.
These findings underscore the critical role of issue-category prediction in enhancing the quality and relevance of generated comments in \tech{}.

\smallskip
\noindent
\textbf{Root Causes Behind Low-Quality Generated Comments.}
Upon our human evaluation, we further conducted a manual analysis of low-quality generated comments (i.e., 69 with average scores below 2) to identify their root causes and provide insights to inform future improvements.
Three root causes are summarized with their frequency (an instance could contain more than one cause):
(1) \textit{Lack of business logic} (22\%).
Many real-world code changes are driven by business requirements rather than functional correctness. 
Without access to business-specific rationale, current methods often misinterpret the intent of changes or flag them as erroneous.
(2) \textit{Confusion over project-specific coding standards} (48\%).
Code edits typically need to follow project-specific coding standards
However, models trained on datasets aggregated from diverse repositories may fail to capture these localized standards (e.g., import ordering), resulting in incorrect or inconsistent comments.
(3) \textit{Limited contextual information} (43\%).
Omitting full function or class definitions restricts the model’s ability to capture semantic dependencies beyond the changed lines, causing it to miss context-dependent issues.
These findings highlight the need for repository-level code review approaches that incorporate business logic, enforce project-specific standards, and utilize broader contextual information to improve the accuracy and relevance of generated comments.

\section{Threats to Validity}
\label{sec:threat}

\textbf{Internal Validity}.~
A potential threat is the data leakage risk in LLMs, as they are trained on open-source projects and may have encountered some test cases during training.
However, our analysis reveals that the LLMs employed in this work exhibit suboptimal performance in zero-shot settings, indicating that their outputs are not solely derived from memorization.
Moreover, given the complexity of the code review task, it is inherently unlikely that LLMs can generate results solely through memorization.
This concern has been similarly acknowledged in other LLM studies~\cite{DBLP:conf/icse/GengWD00JML24,DBLP:conf/icse/NashidSM23}.
Another potential threat lies in the potential positional bias in the fine-tuned critic agent. 
Prior studies~\cite{DBLP:conf/iconip/BaoGPZTCWL24,DBLP:journals/corr/abs-2507-13949} show that while discriminative LLMs can exhibit permutation invariance, fine-tuning does not always eliminate such effects.
To mitigate this, we adopt a structured input format to enforce consistent element mapping, though minor bias may remain.
Finally, to mitigate the threat of the inherent randomness in the LLM inference process, we set the temperature to 0 to stable answers from the LLMs, facilitating reproducibility in future research.

\smallskip
\noindent
\textbf{External Validity}.
The primary threat to external validity is that the dataset was annotated using LLaMA-3.1-70B rather than fully human-labeled, which may introduce annotation errors~\cite{DBLP:journals/corr/abs-2502-03425}.
To mitigate this, all experiments were conducted on the same dataset, ensuring consistent comparisons and minimizing the impact of potential noise.
Another limitation lies in the class imbalance of the test set, where Refactoring accounts for a disproportionately large share. 
This skews overall performance metrics and may introduce bias.
We address this by reporting detailed, category-level results, which highlight \tech{}’s effectiveness across diverse issue types.
Finally, our study fine-tuned only a few representative LLMs with parameter sizes under 14B and did not include proprietary models such as GPT-4o due to resource constraints.
We regard our work as a foundation and remain open to incorporating more advanced and larger models in future studies.

\smallskip
\noindent
\textbf{Construct Validity}. 
The primary threat lies in the human evaluation of comment quality, which may be biased due to inherent subjectivity. To mitigate this, we recruited two participants with substantial experience in Java development, who are independent and not involved in the authorship of this paper. Furthermore, following established practices, we systematically assessed their consistency and comprehension using the inter-rater agreement metric (i.e., FCohen's kappa) during the manual evaluation.

\section{Conclusion}
\label{sec:conclusion}

In this paper, we propose \tech{}, a novel agent-based framework for code review comment generation, addressing the limitations of existing approaches that rely on a single model and overlook the diverse nature of issues in code changes.
Our evaluation on large-scale datasets shows that \tech{} outperforms state-of-the-art baselines in both textual quality metrics and issue category prediction accuracy, underscoring its overall effectiveness. 
Human evaluation further confirms the quality and usefulness of the generated comments.
Moreover, \tech{} strikes a favorable balance between performance and efficiency.
Meanwhile, our work also opens up several promising future directions, including constructing the fine-grained issue typing beyond the current taxonomy, elevating the target of code review from individual code changes to the function or class level, and leveraging external sources (e.g., repository-level documentation or history information) to support more accurate and context-aware code review.

\begin{acks}

This work was supported by the National Natural Science Foundation of China under Grant Nos. 62472310, 62322208. 

\end{acks}

\bibliographystyle{ACM-Reference-Format}
\bibliography{sample-base}










\end{document}